\documentclass[a4paper,fleqn,usenatbib,useAMS]{mnras}

\usepackage{threeparttable}
\usepackage{multirow}

\usepackage{graphicx}	
\usepackage{amsmath}	
\usepackage{amssymb}	
\usepackage{multicol}        
\usepackage{bm}
\usepackage{pdflscape}	

\newcommand{\HI}{\hbox{\rmfamily H\,{\textsc i}}}
\newcommand{\HIsub}{\hbox{{\scriptsize H}\,{\tiny I}}}
\newcommand{\MHI}{\hbox{$M_{\HIsub}$}}
\newcommand{\rhoHI}{\hbox{$\rho_{\HIsub}$}}
\newcommand{\OHI}{\hbox{$\Omega_{\HIsub}$}}
\newcommand{\Msun}{\hbox{${\rm M}_{\odot}$}}
\newcommand{\Lsun}{\hbox{${\rm L}_{\odot}$}}

\defcitealias{Chen:2021a}{Paper I}

\usepackage[T1]{fontenc}
\usepackage{ae,aecompl}

\graphicspath{{./Figs/}}

\usepackage{txfonts}

\title[$\varOmega_{HI}$ Measurement via Stacking]{Measuring Cosmic Density of Neutral Hydrogen via Stacking the DINGO-VLA Data}

\author[Q. Chen et al.]{Qingxiang Chen,$^{1,2}$\thanks{E-mail: chenqingxiangcn@gmail.com}
Martin Meyer,$^{1,3}$
Attila Popping,$^{1,2}$
Lister Staveley-Smith,$^{1,3}$
\newauthor
Julia Bryant,$^{3,4,5}$
Jacinta Delhaize,$^{6}$
B. W. Holwerda,$^{7}$
M. E. Cluver,$^{8,9}$
J. Loveday,$^{10}$
\newauthor
Angel R. Lopez-Sanchez,$^{3,11,12}$
Martin Zwaan,$^{13}$
E. N. Taylor,$^{8}$
A. M. Hopkins,$^{11}$
\newauthor
Angus Wright,$^{14}$
Simon Driver,$^{1,15}$
S. Brough${^{16}}$
\\
$^{1}$The International Centre for Radio Astronomy Research (ICRAR), University of Western Australia, 35 Stirling Hwy, Crawley, WA 6009, Australia\\
$^{2}$ARC Centre of Excellence for All-sky Astrophysics (CAASTRO)\\
$^{3}$ARC Centre of Excellence for All Sky Astrophysics in 3 Dimensions (ASTRO 3D) \\
$^{4}$Australian Astronomical Optics, AAO-USydney, School of Physics, University of Sydney, NSW 2006, Australia\\
$^{5}$Sydney Institute for Astronomy (SIfA), School of Physics, Faculty of Science, The University of Sydney, NSW 2006, Australia\\
$^{6}$Department of Astronomy, University of Cape Town, Private Bag X3, Rondebosch 7701, South Africa\\
$^{7}$Department of Physics and Astronomy, 102 Natural Science Building, University of Louisville, Louisville KY 40292, USA\\
$^{8}$Centre for Astrophysics and Supercomputing, Swinburne University of Technology, John Street, Hawthorn, 3122, Australia\\
$^{9}$Department of Physics and Astronomy, University of the Western Cape, Robert Sobukwe Drive, Bellville, 7535, South Africa\\
$^{10}$Astronomy Centre, University of Sussex, Falmer, Brighton BN1 9QH, UK\\
$^{11}$Australian Astronomical Optics, Macquarie University, 105 Delhi Rd, North Ryde, NSW 2113, Australia\\
$^{12}$Department of Physics and Astronomy, Macquarie University, NSW 2109, Australia\\
$^{13}$ European Southern Observatory, Karl-Schwarzschildstrasse 2, D-85748 Garching bei München, Germany\\
$^{14}$Ruhr-Universität Bochum, Astronomiches Institut, German Center for Cosmological Lensing, Universitätsstr 150, 44780, Bochum, Germany\\
$^{15}$School of Physics \& Astronomy, University of St Andrews, North Haugh, St Andrews KY16 9SS, UK\\
$^{16}$ School of Physics, University of New South Wales, NSW 2052, Australia
}

\date{Last updated 2016 Nov 17}

\pubyear{2019}

\begin{document}
\label{firstpage}
\pagerange{\pageref{firstpage}--\pageref{lastpage}}

\maketitle 

\begin{abstract}
We use the 21 cm emission line data from the DINGO-VLA project to study the atomic hydrogen gas {\HI} of the Universe at redshifts $z<0.1$. Results are obtained using a stacking analysis, combining the {\HI} signals from 3622 galaxies extracted from 267 VLA pointings in the G09 field of the Galaxy and Mass Assembly Survey (GAMA).
Rather than using a traditional one-dimensional spectral stacking method, a three-dimensional cubelet stacking method is used to enable deconvolution and the accurate recovery of average galaxy fluxes from this high-resolution interferometric dataset. By probing down to galactic scales, this experiment also overcomes confusion corrections that have been necessary to include in previous single dish studies.  After stacking and deconvolution, we obtain a $30\sigma$ {\HI} mass measurement from the stacked spectrum, indicating an average {\HI} mass of ${\MHI}=(1.674\pm 0.183)\times 10^{9}~{\Msun}$. The corresponding cosmic density of neutral atomic hydrogen is ${\OHI}=(0.377\pm 0.042)\times 10^{-3}$ at redshift of $z=0.051$. These values are in good agreement with earlier results, implying there is no significant evolution of {\OHI} at lower redshifts.
\end{abstract}




\section{Introduction}\label{section:introduction}

Following decades of effort, the evolution of the cosmic star formation rate has been well measured for redshifts $z<3$. It is now known that the star formation rate density has dropped by more than an order of magnitude since $z \sim 1$ \citep{Lilly:1996,Madau:1996,Hopkins:2004, Hopkins:2006, Hopkins:2008a, Madau:2014}. In contrast, less is known about how the cold gas content of galaxies has evolved during the same cosmic period \citep[e.g.][]{Meyer:2004, Giovanelli:2005, Giovanelli:2015}. Since cold gas is the fuel for future star formation, understanding its availability via observations and theory is essential for the development of a holistic picture of the physics of galaxy formation and evolution.


Much effort has been expended in developing models which relate the physical mechanisms responsible for the accretion and outflow of gas, and the collapse of cold gas clouds in galaxies over cosmic time, and the identification of the major physical mechanisms \citep[e.g.][]{Lagos:2018}. There are many models dealing with different physical mechanisms at a variety of spatial and mass scales, such as the complex gas dynamics of AGN and supernova feedback \cite[e.g.][]{Somerville:2001,Cen:2003,Nagamine:2005,Power:2010,Lagos:2011}.
As such, observational measurements and constraints are becoming even more crucial.

At high redshifts, the damped Lyman-$\alpha$ absorption systems (DLAs) are often used as a tracer of neutral atomic hydrogen gas.
Using spectroscopic data from the Sloan Digital Sky Survey (SDSS), the {\HI} cosmological mass density can be measured at $z > 2$ \citep[e.g.][]{Prochaska:2005,Prochaska:2009}. These results may contain systematic biases due to dust extinction \citep[e.g.][]{Ellison:2001, Jorgenson:2006} and gravitational lensing \citep[e.g.][]{Smette:1997}. At $z<1.6$, Lyman-$\alpha$ enters the ultra-violet regime, and becomes hard to detect using ground-based telescopes. In the local Universe, on the other hand, the preferred method to trace {\HI} is to directly observe the 21-cm emission line of atomic hydrogen. Thanks to large 21-cm emission line blind surveys, the neutral hydrogen mass function and density have been precisely measured in the local Universe at $z\sim 0$ \citep{Zwaan:2005,Martin:2010,Jones:2018}. 
However, beyond the local Universe, direct detection of {\HI} is very challenging due to the relative weakness of the {\HI} signal compared to the sensitivity of existing observing facilities. 
Deep blind surveys, such as the Arecibo Ultra-Deep Survey \citep[AUDS,][]{Hoppmann:2015}, the Blind Ultra-Deep {\HI} Environmental Survey \citep[BUDHIES,][]{Gogate:2020} and the COSMOS {\HI} Large Extragalactic Survey \citep[CHILES,][]{Fernandez:2013, Hess:2019} are able to probe higher redshift {\HI}, but require extremely long integration times. At certain redshifts, where the signal falls outside the protected radio astronomy band, the impact of radio frequency interference (RFI) can also severely limit sensitivity \citep[e.g.][]{Fernandez:2016}.

Thus, there are still considerable uncertainties in our understanding of {\HI} evolution at $z > 0.1$,  and therefore the relationship between {\HI} gas content and the dramatic decrease of the star formation rate density. The next generation of radio telescopes such as the Australian Square Kilometre Array Pathfinder \citep[ASKAP,][]{DeBoer:2009}, the Meer-Karoo Array Telescope \citep[MeerKAT,][]{Jonas:2009} and ultimately the Square Kilometre Array \citep[SKA,][]{Carilli:2004} will likely tackle these problems with their better sensitivity and larger field-of-view. Other than developing these instruments, a technique involving the co-adding of signals from hundreds or thousands of galaxies whose {\HI} signals are too weak to detect directly, has been developed. It was first introduced by \citet{Zwaan:2000a}, and later by \citet{Chengalur:2001a}, probing the gas content of cluster galaxies. Using this technique, the cosmic density of neutral hydrogen can be measured to higher redshifts using single-dish radio telescopes, such as the Parkes telescope, and interferometers such as the Westerbork Synthesis Radio Telescope (WSRT) and the Giant Metrewave Radio Telescope (GMRT) \citep[e.g.][]{Delhaize:2013, Rhee:2013, Hu:2019, Lah:2007, Lah:2009, Rhee:2016, Rhee:2018, Bera:2019,Chowdhury:2020}.

In this paper we develop and apply a new {\HI} stacking method for the Karl G. Jansky Very Large Array (VLA) pathfinder project of the ASKAP Deep Investigation of Neutral Gas Origin survey \citep[DINGO,][]{Meyer:2009b}. This method overcomes observational limitations due to the poor
\textit{uv}-coverage of short observations, and the subsequent non-Gaussian synthesised beam and large sidelobes. Such limitations make it hard to apply the traditional spectral stacking technique when sources are partially resolved. The new method  (\textit{Cubelet Stacking}) solves this problem.
Instead of extracting and stacking spectra, we stack small cubelet cut-outs from the area centered on the known galaxy positions. Then we deconvolve the stacked cubelets using a stacked point spread function and extract a spectrum from this stacked cube. For more information on this stacking technique, readers are recommended to read \citet{Chen:2021a}(hereafter: Paper I).

The paper is organised as follows. Section 2 introduces the optical and radio data used in this work. Section 3 presents our radio data reduction pipeline. In Section 4 we present the sample, summarise the stacking method and show the results. The {\HI} cosmic density is calculated in Section 5, followed by a summary and conclusions in Section 6. We adopt the concordance cosmological parameters of $\Omega_{\Lambda}=0.7$, $\rm \Omega_M=0.3$, and  $\rm H_0=70\ km\ s^{-1}Mpc^{-1}$ and use \citet{Loveday:2012} for the optical luminosity function and density for our analysis.

\section{Data}\label{section:data}

This section introduces both the optical and {\HI} data used in this work. For the analysis of our data we are using an {\HI} stacking technique, as the {\HI} emission lines from most galaxies in our sample are too faint to detect directly. An optical input catalogue providing the positions and redshifts of galaxies is used as a prior to locate the likely position of signal in the {\HI} data.

\subsection{Optical Data}

The optical input catalogues used in this work are from the Galaxy And Mass Assembly survey \citep[GAMA,][]{Driver:2011, Hopkins:2013, Liske:2015}.
GAMA is a multi-wavelength galaxy survey covering several regions of the sky. In this study we only focus on the G09 field covering the RA range between 129$^{\circ}$ and 141$^{\circ}$ and the DEC range from $-2^{\circ}$ to $+3^{\circ}$. One of the core components of GAMA is a large spectroscopic redshift survey carried out with 2dF/AAOmega on the Anglo-Australian Telescope (AAT). This survey observed $\sim$300,000 galaxies with $r<19.8$~mag over $\sim286~\rm{deg}^2$. The G09 catalogue includes coordinates, redshifts, distances, dust extinctions and magnitudes of galaxies from the GAMA 3rd data release \citep{Baldry:2018}. The area of this region is $\sim60~\rm{deg}^2$ and the magnitude limit is $r<19.8$.

The G09 input redshifts used in this paper are from the combined catalog \texttt{SpecALL}, which combined redshifts from the GAMA AAT spectroscopic observations and other publicly available catalogs including SDSS/BOSS DR10 \citep{Ahn:2014a}, 6dF Galaxy Survey \citep{Jones:2009a}, 2dF SDSS LRG and QSO survey \citep{Cannon:2006a, Croom:2009a}, WiggleZ Dark Energy Survey \citep{Parkinson:2012a} and Updated Zwicky Catalog \citep{Falco:1999a}. The redshift quality is encoded by $NQ$ ($1- 4$), ranked from failure to most confident. Throughout this paper we select galaxies with secure redshifts with $NQ>2$. The overall completeness of secure redshifts in G09 is 98.48\% at $r<19.8$. Interested readers are referred to \citet{Liske:2015} and \citet{Baldry:2014a} for extensive discussions on how these redshifts are measured and categorized. Also, in order to avoid confusion with star-like objects, galaxies with a heliocentric redshift below 0.002 are not considered.
We use two different types of GAMA redshifts \citep{Baldry:2018}: heliocentric redshifts are used to determine the location of galaxies in the radio data. To calculate luminosity distances, we use the CMB redshifts corrected for the local flow model of \cite{Tonry:2000}. The redshift error is only 27 $\rm km\ s^{-1}$ and should not affect our stacking analysis. It should slightly smear out flux along the frequency axis of our stacked cubes, but conserve it. Therefore we do not consider redshift errors in this work.

\begin{figure}
    \includegraphics[width=\columnwidth]{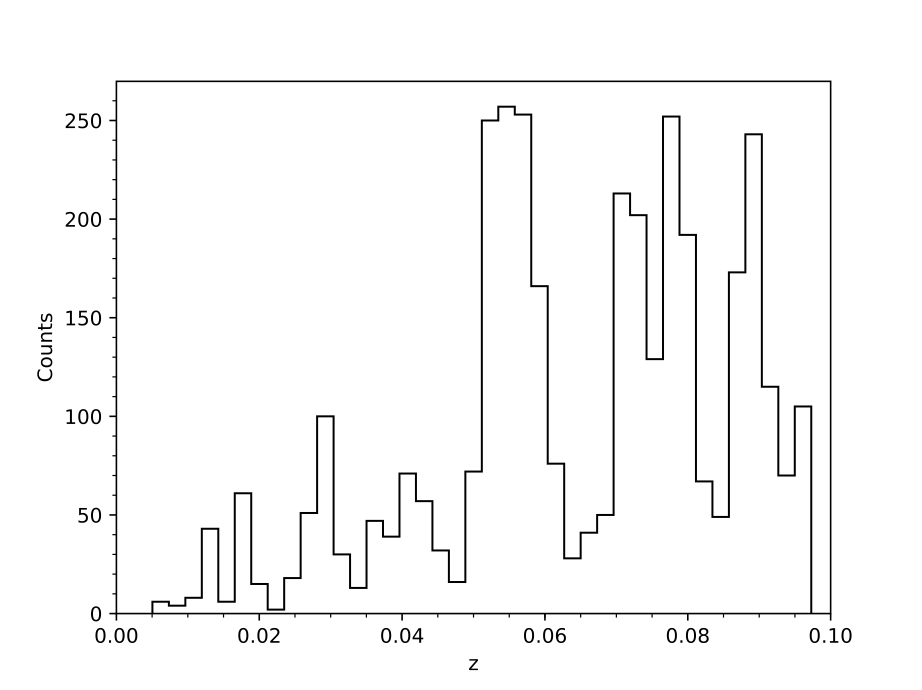}
    \caption{The redshift distribution of the 3622 galaxies with $z<0.1$ stacked in this work. These are CMB redshifts corrected for local flow. 
    }
    \label{fig:z_distr_3622sample}
\end{figure}

The $u, g, r, i, z$ luminosity functions and densities of the G09 galaxies are provided in \citet{Loveday:2012}. For this study, we only use the luminosity function in the $r$-band as it has the highest accuracy. All magnitudes in the catalogues were adjusted for effects due to dust extinction and $k$-correction.  
For this paper, we limit the redshift to the RFI-free redshift range $z<0.1$. A later study will examine data at higher redshifts. We match the optical data with the sky area of VLA data, resulting in a total galaxy sample of 7146. After considering further observational constraints, selecting only galaxies with positions within the VLA primary beam, and selecting only galaxies which have data extending $\pm 2$ MHz from the redshifted {\HI} frequency, the number of galaxies in the sample decreases to 3622. Fig.~\ref{fig:z_distr_3622sample} shows the redshift distribution of the total sample. This galaxy sample was then used to generate 5442 cubelets to co-add, larger than the number of sample galaxies due to some galaxies appearing in multiple pointings (see Fig.~\ref{fig:vla_pointings}).

\subsection{Radio Observations and Data Reduction}\label{section:radio_data}

The radio data is obtained from the DINGO-VLA project observed in the 2014B and 2016A semester with the Jansky Very Large Array (VLA). The observations were taken in the C or CnB configurations.

\begin{figure*}
    \includegraphics[width=\textwidth]{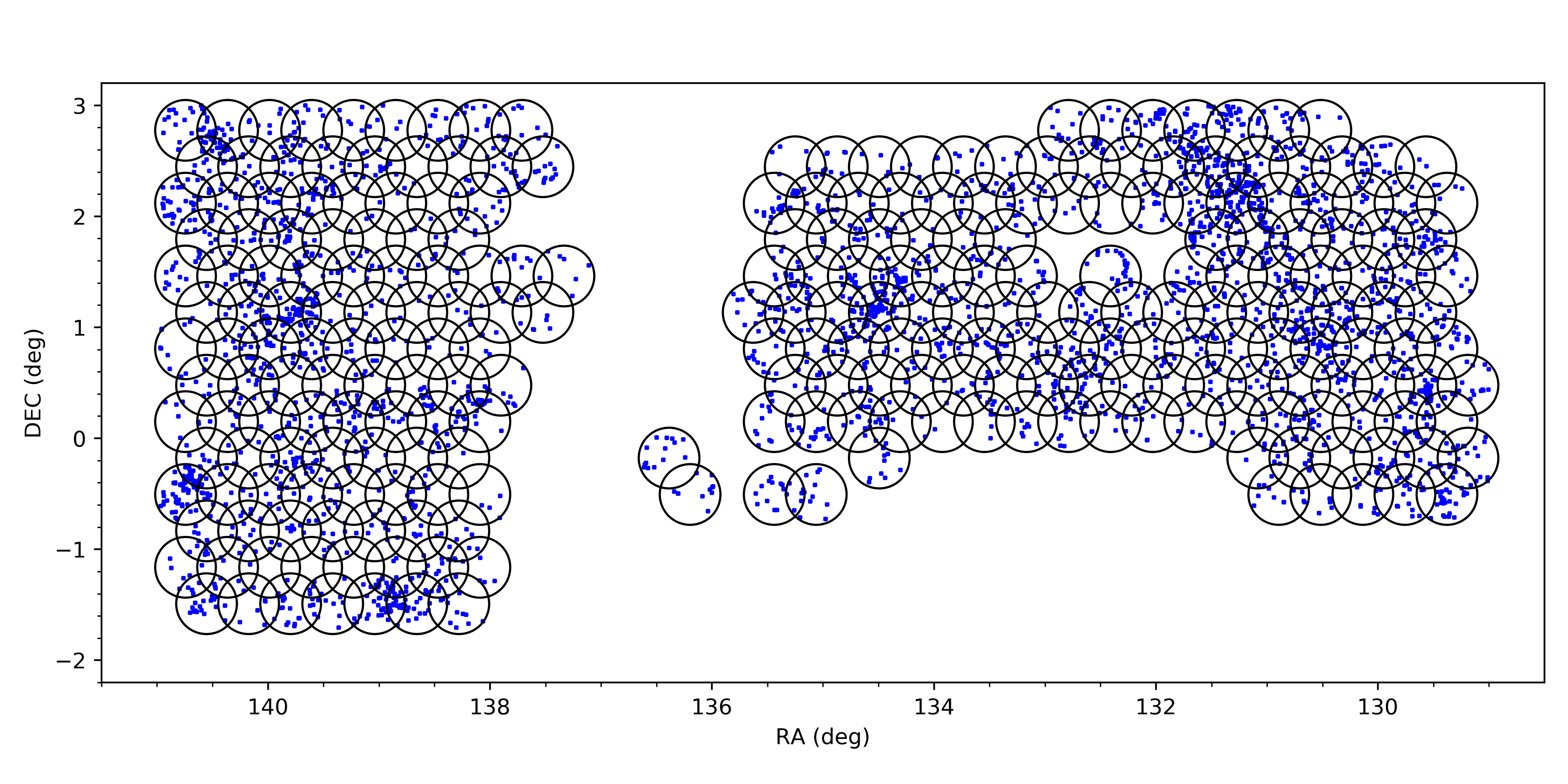}
    \caption{
    The mosaic of pointings of the DINGO-VLA project (circles) from which the radio data used for this work were obtained. The total number of pointings is 267. Each circle indicates one VLA primary beam with a radius of 16.4 arcmin. This approximates the full-width at half power (FWHP) at the frequency corresponding to redshift 0.1. The blue points within circles indicate the positions of the 3622 GAMA galaxies considered in this paper.
    }
    \label{fig:vla_pointings}
\end{figure*}

The DINGO-VLA observations include three target pointings and two calibrators within each two-hour scheduling block. The sky positions of the different target pointings are shown in Fig.~\ref{fig:vla_pointings}.
The complete area of 276 pointings consists of 92 2-hr observing units.
In each observing unit the flux density and bandpass calibrator 3C~138 is observed first, followed by the phase calibrator, then the three target pointings. The observation sequence in one observing unit is: flux calibrator, 4$\times$ (phase calibrator $\to$ pointing 1 $\to$ 2 $\to$ 3), phase calibrator. Observations of pointings are broken into small intervals to maximize the UV coverage. Each pointing target was observed for a total of 28 mins. The full width half power (FWHP) circle radius of VLA primary beam at 1.4~GHz is $\sim31\arcmin$, resulting in $\sim38$~deg$^2$ total survey area.

\begin{figure*}
    \includegraphics[width=\textwidth]{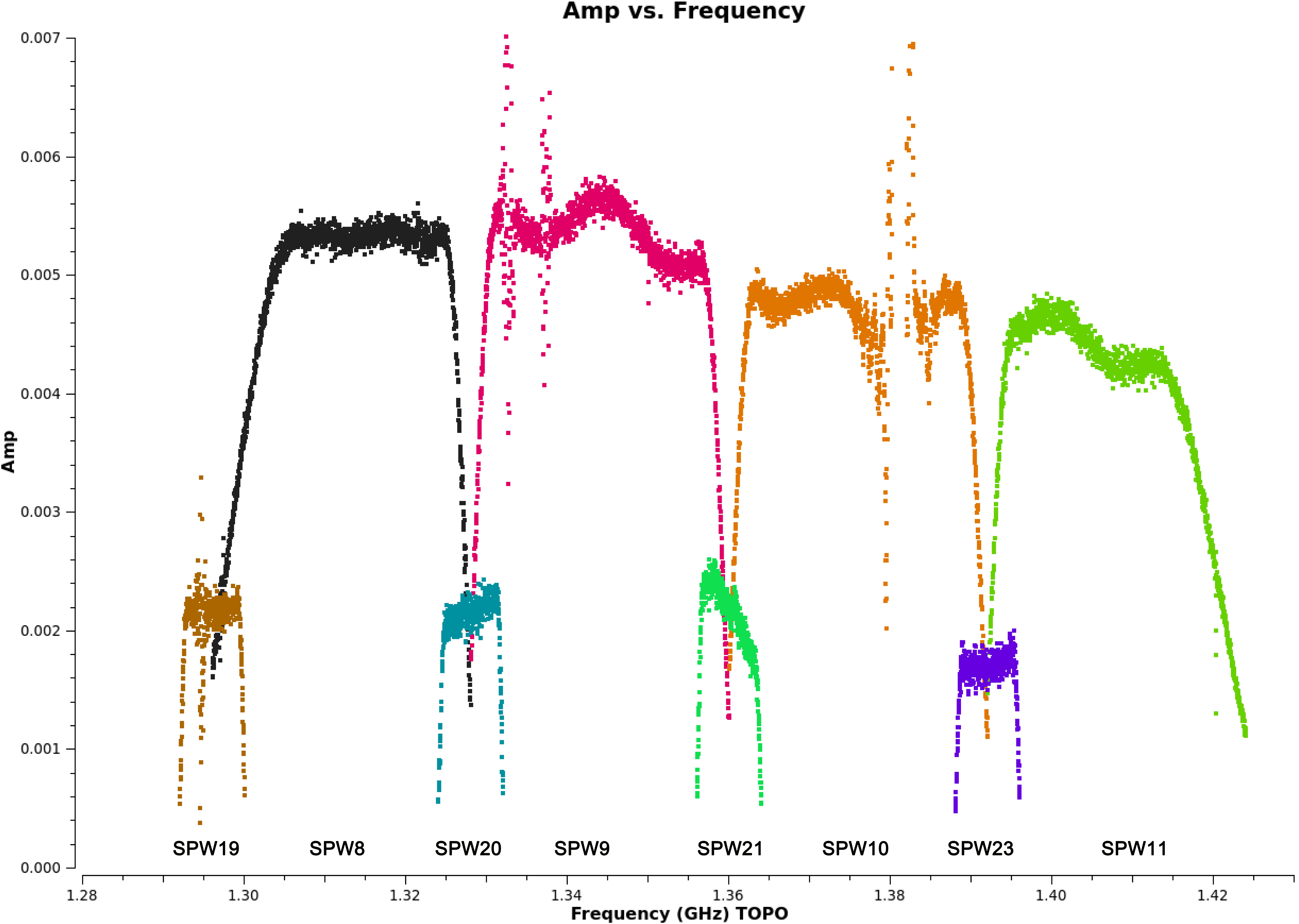}
    \caption{A plot of amplitude against frequency showing the layout of four wide 32-MHz spectral windows of the VLA correlator used in this analysis (\textit{spw8}, \textit{spw9}, \textit{spw10}, \textit{spw11}), and the four narrow 8-MHz spectral windows (\textit{spw19}, \textit{spw20}, \textit{spw21}, \textit{spw23}) positioned to fill the subsequent gaps. Here we plot the bandpass calibrator raw data from a randomly chosen observing unit after averaging along polarisation, time and uv-distance axes. The channel width in the wide and narrow windows is 20.833 and 15.625 kHz, respectively.}
    \label{fig:vla_spws}
\end{figure*}

The correlator of the VLA splits the bandwidth into spectral windows. In this paper we consider 
the 4 spectral windows which cover the frequency range from 1296 to 1424~MHz, denoted as \textit{spw8}, \textit{spw9}, \textit{spw10}, \textit{spw11}. Each \textit{spw} spans 32 MHz, with a channel width of 20.833~kHz. Both edges of each spectral window are noisy because of the bandpass response. In order to deal with this problem, another 4 spectral windows are overlaid across the boundaries of the \textit{spw}s. These four narrow spectral windows are denoted  \textit{spw19}, \textit{spw20}, \textit{spw21}, \textit{spw23}. Each of them spans 8~MHz and their channel width is 15.625~kHz. We show the 8 spectral windows' frequency coverage in Fig.~\ref{fig:vla_spws}. The frequency range after adding the 4 narrow spectral windows becomes 1292 to 1424~MHz. This strategy significantly improves the data quality and ensures a more uniform rms level over the full frequency range.

\begin{figure*}
    \includegraphics[width=\textwidth]{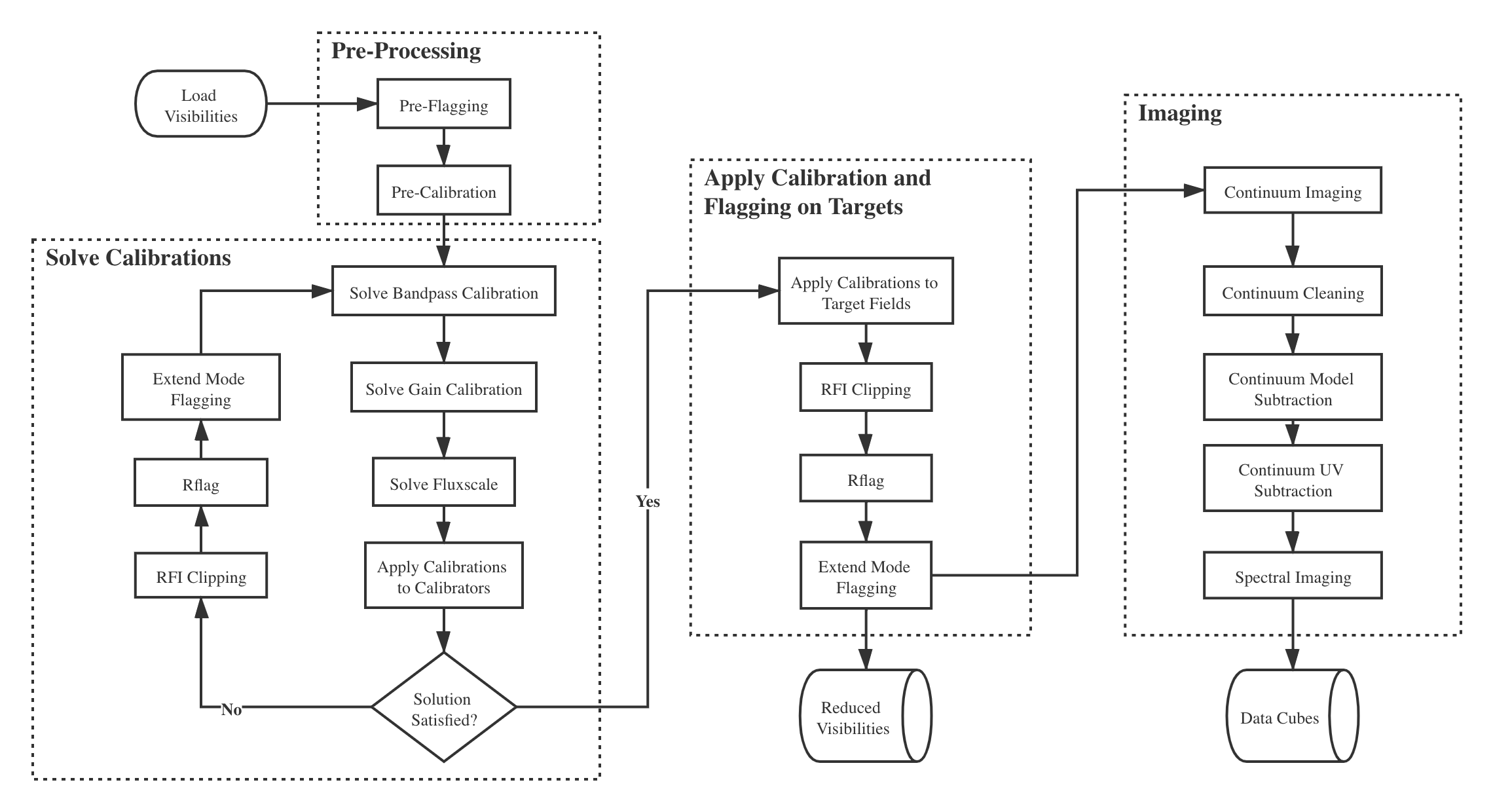}
    \caption{An illustrative flow-chart of the data reduction procedures for DINGO-VLA. A more detailed description of the pipeline is in Section~\ref{section:radio_data}.}
    \label{fig:flowchart}
\end{figure*}

We developed a data reduction pipeline for DINGO-VLA based on standard tasks in Common Astronomy Software Applications \citep[CASA,][]{McMullin:2007}. The pipeline was tuned specifically for the the data properties and the large data volume (raw data in  measurement set format is larger than 16 TB). Scripts were developed to implement the pipeline in a cluster environment. An outline of the data reduction procedure is in Fig.~\ref{fig:flowchart}. It includes pre-processing, visibility processing (flagging and calibration), and an imaging step.

First, a pre-processing step is carried out. This includes application of online flags (data taken when not on source or when there were sub-reflector issues), flagging of zero-amplitude data, flagging of auto-correlations, flagging of shadowed antennas, first scan flagging (telescope setup), and so-called quack flagging (the first 10-15s of data are not useful). In addition, we correct for antenna position errors and antenna shape variation (task: \textit{gencal}).

The main bandpass and flux density calibrator is 3C~138. This is used to to solve for amplitude and phase variations as a function of frequency for all antennas. The second calibrator is used to correct for the complex gain variations (i.e., amplitude and phase) as a function of time. 
Due to some frequency ranges being affected by RFI, calibration and flagging was applied iteratively.
A 2-D automated algorithm (rFlag) was used to flag 5-$\sigma$ outliers in the time-frequency plane. Strong RFI signals (>20~Jy) were flagged. Channels and time ranges were completely flagged where more than 80\%  and 50\% of the time or frequency series were flagged, respectively.
The calibration procedure was then repeated for more accurate solutions.

We then apply the calibration solutions (bandpass, gain and flux scale) to the three target fields. Again, we use the rFlag algorithm to flag bad data on the 2-D plane. To assist rFlag, we apply a 3-$\sigma$ clip based on the standard deviation of amplitude across the spectral window beforehand. After rFlag we carry out an extend mode flagging: completely flag target data in each channel / time chunk where more than 50\% has already been flagged on the 2-D plane; if there are more than 4 surrounding data points flagged on the 2-D plane, this data point is further flagged.

Finally we carry out the imaging processes. The weighting scheme used is `robust' \citep{Briggs:1995} with a robust parameter of 0.8. This value was found to be the optimal compromise between resolution and sensitivity for DINGO-VLA. For every target field, we first obtain the continuum image by carrying out multi-frequency synthesis (MFS) mode imaging \citep{Rau:2011}. The size of each continuum image is 2048 pixels $ \times$ 2048 pixels, with the pixel size being $2''\times 2''$. We calculate the rms over all the pixels in the dirty image and deconvolve down to a threshold of 5 times this value. Then we subtract the MFS {\sc clean} source model from the visibilities, and further subtract the continuum residual by directly fitting the real and imaginary visibility data with linear functions in the frequency domain. The four narrow spectral windows (\textit{spw19}, \textit{spw20}, \textit{spw21}, \textit{spw23}) have channel widths of 15.625~kHz, while the four wide spectral windows (\textit{spw8}, \textit{spw9}, \textit{spw10}, \textit{spw11}) have channel widths of 20.833~kHz. We average 4 and 3 channels for narrow and wide spectral windows respectively, to obtain the same channel width (62.5~kHz). We use the re-binned data for the subsequent analysis. Dirty data cubes are made for each spectral window, and concatenated, discarding the duplicate 1~MHz chunks at both edges of each wide spectral window. Every field then has a single data cube consisting of 1024~pixels $\times$ 1024~pixels $\times$ 2048~channels. The pixel size and channel width is $2''\times2''$ and 62.5~kHz respectively.

\begin{figure}
    \includegraphics[width=\columnwidth]{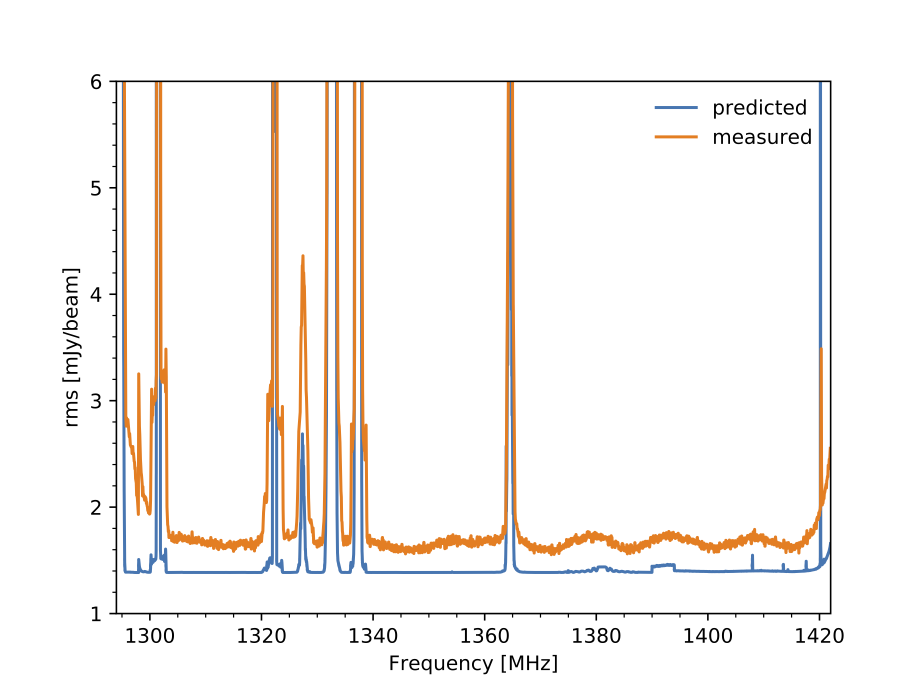}
    \caption{The measured and predicted rms  as a function of frequency for a typical field. 
    The orange line is the rms level directly measured from the reduced data cube. The blue line is the predicted rms level. The latter is calculated by scaling the prediction from the VLA online Exposure calculator, taking into account the flagging fraction. The channel width in this measurement is 62.5~kHz. }
    \label{fig:rfi_ratio}
\end{figure}

For every spectral window we chop 2~MHz at both edges so that the spectral windows are aligned. This also changes the overall frequency range for science use. The final visibility data after reduction covers 128 MHz from 1294 to 1422 MHz (more details can be found in next section).
The predicted and measured noise behaviour as a function of frequency for a typical field is shown in Fig.~\ref{fig:rfi_ratio}. The measured rms is obtained directly from the final data cube, and the predicted rms uses the VLA online exposure calculator\footnote{https://obs.vla.nrao.edu/ect.}, taking into account the flagging ratio, the spectral resolution, the configuration, weighting scheme and integration time.
The average measured rms is 1.65 mJy~beam$^{-1}$, which is within 20\% of the predicted value.

\section{Cubelet Stacking}\label{section:stacking}

As discussed in Paper I, traditional stacking methods are not well-suited to interferometric data. The equatorial location of the observations and short integration times lead to significant power in the wings of the PSF, or dirty beam.  Combined with the extended nature of some input sources compared to the size of the PSF, this has the potential to lead to significant flux errors. We therefore use the \textit{Cubelet Stacking} technique, first extracting and stacking sub-cubes centred on the known galaxy redshifts and positions, then deconvolving the stacked cube.  We then analyse {\HI} mass and density using the spectrum extracted from the clean cube. For more details of the stacking method please refer to Paper I.

\subsection{Stacking Sample Selection}

A number of selection criteria are applied in extracting the \HI\ data cubelets from the reduced VLA spectral-line observations, as detailed below:

\begin{enumerate}
 \setlength\itemsep{1em}
  \item Only GAMA sources that fall within the full width half maximum (FWHM) of the primary beam for each pointing are extracted.  
  \item Sources for which we cannot extract a 200~pixels $\times$ 200~pixels region ($400\arcsec \times 400\arcsec$), centred on the position of the GAMA galaxy, are omitted.  
  \item Each cubelet should be at least 4~MHz wide along the frequency axis, centred on the frequency of the 21-cm line (1420.406~MHz), after aligning the cubelet to the rest frame. This is to avoid spectra with only partial frequency information. Due to the observed frequency range of 1294 to 1422~MHz, this excludes galaxies with redshifts $z>0.095$. There are some completely flagged channels within the central 4~MHz for some of the cubelets. They are included in this sample, but given zero weight in the stacking process.
 \item The redshift quality flag (NQ) in the GAMA optical catalogues is larger than 2.
\end{enumerate}

Applying the above criteria yields a sample of 3622 galaxies for stacking. Due to some galaxies being covered in adjacent pointings, there are a total of 5442 cubelets. Fig.~\ref{fig:z_distr_3622sample} shows the redshift distribution of the sample.

\subsection{Stacking Methodology}\label{subsec:method}

Following extraction of each $400\arcsec\times400\arcsec\times4$ MHz    cubelet centred on the optical position and redshift from GAMA, a corresponding 200 $\times$ 200 pixel PSF cubelet is also extracted from the centre of the PSF cube generated by the pipeline for each observation.  These are used for stacking as described below.

\begin{enumerate}
 \setlength\itemsep{1em}

 \item \textit{Primary Beam Correction.} For each image cubelet, we calculate the distance to the pointing centre of the observation, and calculate the primary beam correction factor for this cubelet based on its position using the VLA primary beam response:

\begin{equation} \label{PB}
f = a_0+a_1 X^2 + a_2 X^4 + a_3 X^6,
\end{equation}
where $X=\nu \theta$ where $\theta$ is the angle between the source and pointing centre in arcmin, and $\nu$ is the observation frequency in GHz. The four coefficients are provided in \citet{Perley:2016}: 1.000, $-1.428\times 10^{-3}$, $7.62\times 10^{-7}$, $-1.54\times 10^{-10}$ at 1296 MHz; 1.000, $-1.449\times 10^{-3}$, $8.02\times 10^{-7}$, $-1.74\times 10^{-10}$ at 1360 MHz and 1.000, $-1.462\times 10^{-3}$, $8.23\times 10^{-7}$, $-1.83\times 10^{-10}$ at 1424 MHz. Responses at intermediate frequencies are derived by linearly interpolation. 
A primary beam correction is applied to each image cubelet by dividing pixel values with the primary beam response $f$. To limit the point spread function variance and enhance the chances of successful deconvolution, we apply the same primary beam correction to all pixels in the cubelet.


\item \textit{Blueshifting to the rest frame.} Each image cubelet and corresponding PSF cubelet is shifted to rest frame using the optical  redshift:
\begin{equation} \label{eq:blueshift}
\nu_{\rm rest}=\nu_{\rm obs}(1+z),
\end{equation}
where $\nu_{\rm rest}$ and $\nu_{\rm obs}$ are the rest frame and observation frame frequency, respectively.  To ensure flux is conserved after blueshifting, the image cubelet pixel values $S_{ij}(\nu_{\rm obs})$ are also scaled with redshift to $S_{ij}(\nu_{\rm rest})$ as follows:

\begin{equation} \label{eq:blueshift_flux}
S_{ij}(\nu_{\rm rest})=\dfrac{S_{ij}(\nu_{\rm obs})}{(1+z)}
\end{equation}

\item \textit{Conversion to mass density.} For each image cubelet we convert every pixel from flux units (Jy~beam$^{-1}$) to {\HI} mass ({\Msun}~beam$^{-1}$) using \citep{Meyer:2017}:
\begin{equation} \label{HI_mass}
\dfrac{{\MHI^{'}(\nu)}}{\rm \Msun~beam^{-1}~channel^{-1}}=49.7\left(\dfrac{D_L}{\rm Mpc}\right)^2\left(\dfrac{S(\nu)\cdot\Delta f}{\rm Jy~beam^{-1} Hz}\right)
\end{equation}
In this equation, $D_L$ is the luminosity distance calculated using the cosmological parameters given in Section~\ref{section:introduction}, and $\Delta f$ is the frequency channel spacing after blueshifting.
Note that after this step, the image cubelet pixels are in units of {\Msun}~beam$^{-1}$~channel$^{-1}$.

\item \textit{Frequency interpolation.} We interpolate all the image and PSF cubelets to have the same rest-frame channelisation of 160 $\times$ 62.5~kHz (10~MHz), each channel being 62.5~kHz (13.2 km~s$^{-1}$) in order to facilitate combination. This provides enough channels for baseline fitting and subtraction after extracting the spectrum.





 \item \textit{Weighting.} For each image cubelet and corresponding PSF cubelet, we calculate the weight in each channel from:
\begin{equation} \label{eq:wt_chan}
w_i=\sigma_i^{-2}D_L^\gamma ,
\end{equation}
where $\sigma_i$ is the rms for each channel, and $D_L$ is the luminosity distance. 
 The rms noise is calculated from the corresponding channel in the original data cube.
 The distance index $\gamma (\le 0)$ is introduced as a parameter that can be adjusted to emphasise nearby galaxies (for good S/N ratio, for example), or to emphasise distant galaxies (for lower cosmic variance, for example). 
 
 \item \textit{Stacking image and PSF cubelets.} We co-add the image and PSF cubelets channel-by-channel using: 
 \begin{equation} \label{eq:stk_chans}
M_{ij}(\nu_{\rm grid})=\dfrac{\sum_{l} M_{ijl}(\nu_{\rm grid})w_{l}(\nu_{\rm grid})}{\sum_{l} w_{l}(\nu_{\rm grid})},
\end{equation}
%
%
and
\begin{equation} \label{eq:stk_chans_psf}
P_{ij}(\nu_{\rm grid})=\dfrac{\sum_{l} P_{ijl}(\nu_{\rm grid})w_{l}(\nu_{\rm grid})}{\sum_{l} w_{l}(\nu_{\rm grid})}
\end{equation}

For each cubelet, empty channels are given zero weight, but normal channels are not affected.


 \item \textit{Deconvolution.} We deconvolve the stacked image cubelets using the stacked PSF cubelet. Note that this is purely done in the image domain using the \textit{CASA} task \textit{deconvolve}. Following Paper I we implement a H\"{o}gbom {\sc clean} with a multi-scale parameter of [6~pix, 12~pix, 18~pix], a central circular cleaning mask of radius $20\arcsec$, and {\sc clean} down to 1-$\sigma$, where $\sigma$ is calculated from all pixels, excluding the central 33 channels (3.0625~MHz) where we expect {\HI}.
The deconvolved image is restored with a two dimensional Gaussian function fitted to the stacked PSF cubelet.


 \item \textit{Mass spectrum.} We extract the mass spectrum from the restored image using a circular aperture of radius $20^{''}$ (denoted as $A$) using: 
\begin{equation} \label{eq:mass_spectrum}
\dfrac{M_{A}(\nu_{\rm grid})}{\Msun~\rm { beam}^{-1}channel^{-1}} = \dfrac{\sum_{A}M_{ij}(\nu_{\rm grid})}{\sum_{A}G_{ij}(\nu_{\rm grid})} ,
\end{equation}
where $G_{ij}$ represents the pixel values from the Gaussian restoring beam cubelet.

\end{enumerate}

Whilst it would be possible, we decided not to stack our galaxy sample on the same physical scale, but rather chose to stack based on the same angular resolution (i.e. the 2$^{''}\times$2$^{''}$ being used for each pixel.). As we are only interested in recovering total {\HI} mass in this work, stacking on the same physical scale is not required, and stacking on the same angular scale will limit the point spread function variance and hence enhance the chances of successful deconvolution.

\subsection{Results and Error Analysis}\label{s:resultserrors}

\begin{figure*}
    \includegraphics[width=\textwidth]{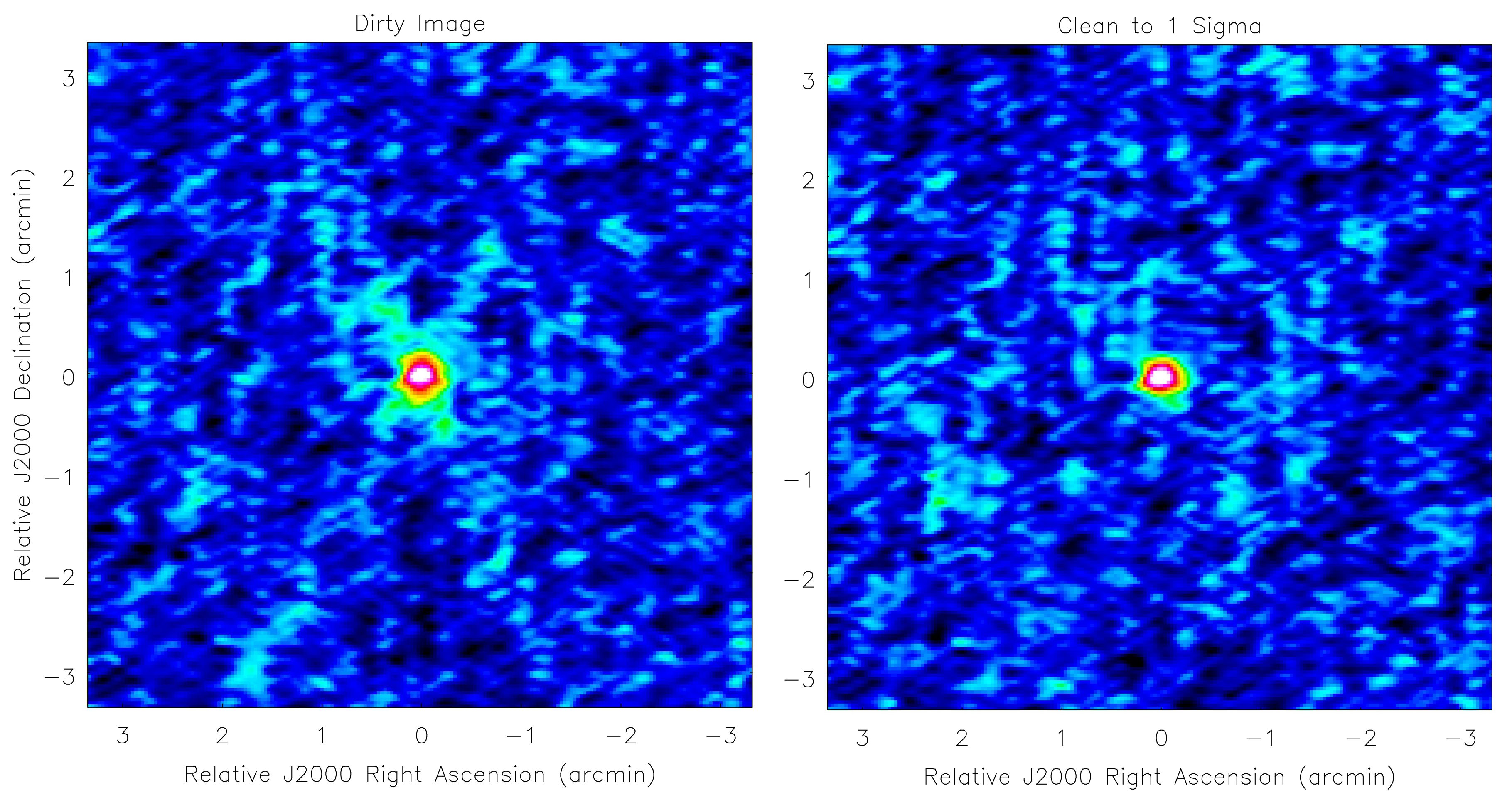}
    \caption{Example images of the stacked {\HI} signal, integrated across a bandwidth of 3.0625~MHz, before and after deconvolution. Left: the moment 0 stacked image of 5442 cubelets before deconvolution. Right: the same image after deconvolving the stacked cubelet to a 1-$\sigma$ {\sc clean} threshold. Sidelobes are effectively suppressed by deconvolution. A weight parameter $\gamma=-1$ is used.}
    \label{fig:deconvolve_comp}
\end{figure*}

Fig.~\ref{fig:deconvolve_comp} shows the stacked image before and after the deconvolution for the case of distance index $\gamma=-1$. The restoring beam, derived from a fit to the stacked PSF cubelet, is $16\arcsec\times12\arcsec$. The image quality is considerably improved after deconvolution. As indicated in Paper I, the good image quality results from the better uv coverage achieved in the PSF stacking process (due to increased hour angle coverage), and the fact that there exists a strong source suitable for deconvolution. This is not possible before stacking.

\begin{figure*}
    \includegraphics[width=\textwidth]{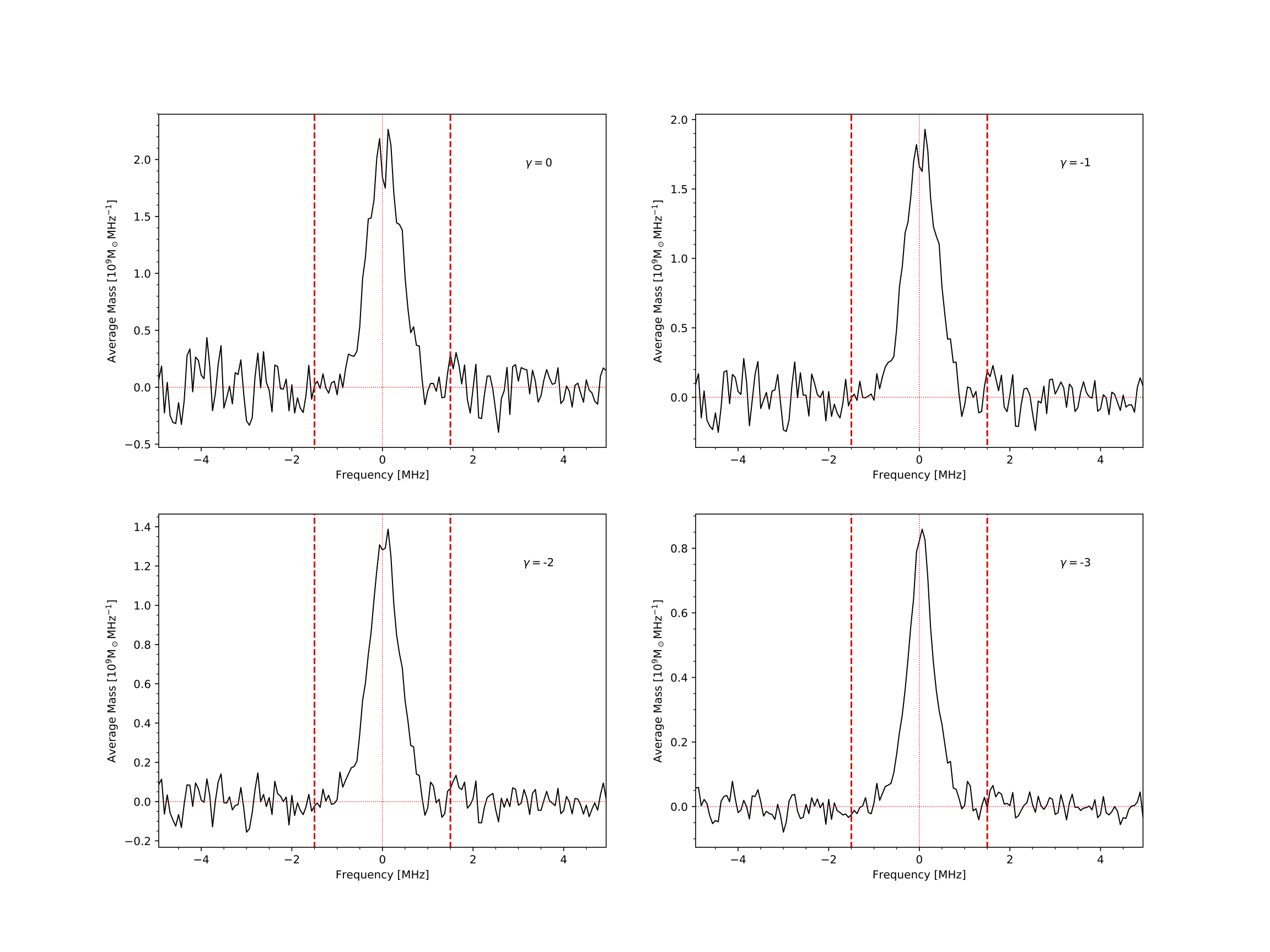}
    \caption{Stacked mass spectra extracted from the deconvolved cubelets for four different values of the weighting parameter, $\gamma$. The aperture radius used is $R=20\arcsec$. The horizontal axis is centred on the {\HI} rest-frame frequency. The vertical axis is the stacked {\HI} mass per frequency interval. The two dashed vertical lines enclose the central 33 channels (3.0625~MHz) which are used for deriving the total {\HI} mass. The average mass decreases for lower values of $\gamma$, consistent with expectation for a magnitude/flux-limited sample when bright distant galaxies are down-weighted.}
    \label{fig:spectrum-mass}
\end{figure*}

Mass spectra are extracted using Eq.~\ref{eq:mass_spectrum}, and a constant baseline is then fit to the spectra. The final stacked, deconvolved and extracted spectra for $\gamma=0,-1,-2,-3$ are shown in Fig.~\ref{fig:spectrum-mass}. 
An integration interval of 49 channels centered on {\HI} restframe frequency is used to derive the total {\HI} mass from the stacked spectra. It equals to 3.0625~MHz and is conservatively wide enough to enclose the stacked signals.
Errors in total {\HI} mass are obtained from jackknife sampling \citep{Efron:1982}, using 20 jackknife re-samples of the 5442 cubelets for each value of $\gamma$, followed by a repeat of the stacking and deconvolution procedure. 





\begin{figure}
    \includegraphics[width=\columnwidth]{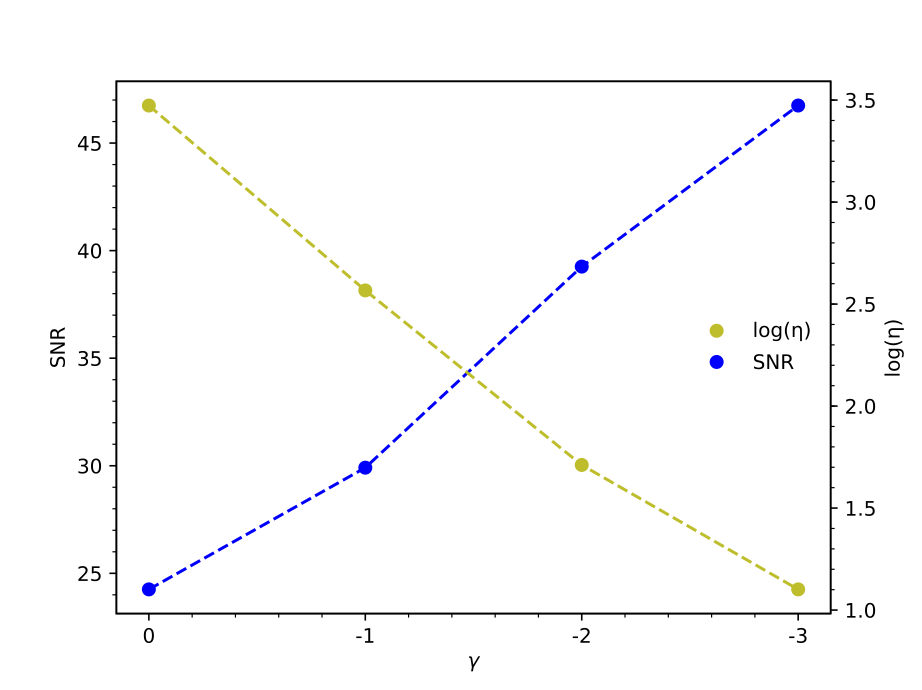}
    \caption{S/N ratio for the stacked {\HI} mass measurement and the effective sample size $\eta$, as a function of the distance weighting factor $\gamma$. With lower values of $\gamma$, the S/N ratio increases but the effect of low sample size (and cosmic variance) becomes more significant. The effective sample size is 3622 without distance weighting ($\eta=0$), but drops to $\sim 20$ for $\gamma=-3$. 
    }
    \label{fig:snr_eta}
\end{figure}

The re-sampled spectra are fit with Gaussian functions, which are used to derive jackknife errors, and peak and integrated signal-to-noise ratios, respectively. Table~\ref{tab:HI_table} summarises our results. The average {\HI} mass decreases with decreasing $\gamma$. This is expected, as smaller values of $\gamma$ highlight nearby sub-samples, which include relatively low gas mass galaxies compared to distant sub-samples. The integrated S/N ratio from Table~\ref{tab:HI_table} is plotted as a function of $\gamma$ in Fig.~\ref{fig:snr_eta}. S/N ratio increases from 24.3 to 46.7 for values from $\gamma=0$ to $\gamma=-3$, respectively. 

However, the effective sample size -- defined as $\eta=\sum_{l} w_l / \max(w)$ -- drops dramatically when nearby galaxies are highly weighted. With no distance weighting ($\gamma=0$), the galaxy sample size is 3622 (only image noise in the observed frame is taken into account), whereas $\gamma=-3$ results in an effective sample size of $\eta\sim20$, representing a significant increase in Poisson error and cosmic variance. 
Values of $\gamma=0$ (no distant-dependent weighting, large cosmic volume) and $\gamma=-1$ (higher S/N ratio) are commonly seen in the literature \citep{Delhaize:2013, Rhee:2013, Rhee:2016, Rhee:2018, Hu:2019}.
 


\begin{table*}
	\centering
	\caption{Final results after stacking, deconvolution and jackknife error analysis. An aperture radius of $20\arcsec$  is used to extract masses. Column~1 is the weighting index; columns~2-4 are the weighted averaged redshift, \textit{r}-band luminosity and measured {\HI} mass, respectively; columns~5 and 6 are the integrated and peak S/N ratios of the extracted mass spectrum; column~7 is the completeness factor for cosmic {\HI} density; column~8 is the derived H\,\textsc{i} cosmic density.}
	\label{tab:HI_table}
	\begin{threeparttable}
        \begin{tabular}{rccccccc}
                \hline
               $\gamma$  & $ \left \langle z \right \rangle$ & $\left \langle L_r \right \rangle$ & $\left \langle {\MHI} \right \rangle$ & Integrated SNR\tnote &  Peak SNR  & $f$ & ${\OHI}$ \\
                   &  & $ (10^9 {\Lsun})$  & $(10^9 {\Msun})$ & &  &  & $(10^{-3})$ \\
                 (1) & (2) & (3) & (4) & (5) & (6) & (7) & (8) \\
               \hline
               0 & 0.063                      & 6.737                          & 1.981 $\pm$ 0.241 & 24.3 & 11.7 & 1.442        & 0.382 $\pm$ 0.045       \\
              $-1$ & 0.051                      & 5.607                          & 1.674 $\pm$ 0.183 & 29.9 & 14.6 & 1.400         & 0.377 $\pm$ 0.042      \\
              $-2$ & 0.035                      & 3.718                          & 1.163 $\pm$ 0.129 & 39.3 & 20.1 & 1.290         & 0.364 $\pm$ 0.041      \\
              $-3$ & 0.020                      & 1.851                          & 0.626 $\pm$ 0.110 & 46.7 & 27.1 & 1.075         & 0.329 $\pm$ 0.058       \\ \hline
        \end{tabular}
    \end{threeparttable}
\end{table*}




Cosmic variance errors are further investigated by investigating four of the GAMA survey regions: G02, G09, G12 and G15 \citep{Baldry:2018}. These four fields have $r$-band magnitude limits of 19.8, 19.0, 19.0, 19.8, respectively and sky areas of $\sim 60$~deg$^2$ (except G02, with $\sim 55.7$~deg$^2$). Our DINGO-VLA tiles have total coverage of $\sim 38$~deg$^2$, or a comoving volume of $\sim 1.620\times 10^7$~Mpc$^3$. Within each of the 4 GAMA fields we extract two rectangular regions, as defined in Table~\ref{tab:8mock}. In each of these 8 sub-fields we define the GAMA sub-samples with $NQ>2$, $z<0.095$ and $r<19.0$. The resultant sample size is also listed in Table~\ref{tab:8mock}. We calculate total weights $\sum D_L^\gamma$ for each region. The cosmic variance (the ratio of the standard deviation to the average value for the 8 sub-fields) is: 29\%, 30\%, 53\%, and 80\% for $\gamma=0,-1,-2,-3$, respectively. This again shows the increasing effect of cosmic variance with decreasing values of $\gamma$. 

For our final analysis, we use $\gamma=-1$. As shown above, the cosmic variance is likely to be very similar to that for $\gamma=0$, but the S/N ratio of the mass estimate is improved by $\sim23\%$ (Table~\ref{tab:HI_table}).
With this value, the average mass is ${\MHI}=(1.674\pm0.183)\times 10^9 {\Msun}$. However, to continue to judge likely effects arising from cosmic variance and other systematic effects, we continue to explore variation with $\gamma$.

\begin{table}
	\centering
	\caption{Eight sub-fields defined from GAMA survey regions in order to investigate cosmic variance. Two 38 deg$^2$ regions are chosen from each of the G02, G09, G12, and G15 regions (G23 does not have the similar $r$-band completeness).}
	\label{tab:8mock}
        \begin{tabular}{lccc}
                \hline
               Region & RA & Declination & Galaxy counts \\
                       &   (deg)  &  (deg)  &  \\
               \hline
               G02-1 & 30.2 $\sim$ 38.8   & $-10.25$ $\sim$ $-5.83$ & 1733  \\
               G02-2 & 30.2 $\sim$ 38.8   & $-8.14$ $\sim$ $-3.72$  & 3016  \\
               G09-1 & 129 $\sim$ 136.6   & $-2$ $\sim$ 3         & 3460  \\
               G09-2 & 133.4 $\sim$ 141   & $-2$ $\sim$ 3         & 3130  \\
               G12-1 & 174 $\sim$ 181.6   & $-3$ $\sim$ 2         & 5203  \\
               G12-2 & 178.4 $\sim$ 186   & $-3$ $\sim$ 2         & 4853  \\
               G15-1 & 211.5 $\sim$ 219.1 & $-2$ $\sim$ 3         & 4885  \\
               G15-2 & 215.9 $\sim$ 223.5 & $-2$ $\sim$ 3         & 3844  \\
               \hline
        \end{tabular}
\end{table}

\section{Cosmic {\HI} Density {\OHI}}\label{section:density}

\subsection{Method 1: completeness correction}


The {\HI} mass derived above can now be used to derive the corresponding {\HI} mass density ({\rhoHI}). A straightforward way is to divide the {\MHI} with the volume in which the observations were carried out. 
Unfortunately, our parent sample is not volume-limited.
We thus adopt a method previously used to measure {\HI} density from {\HI} stacking experiments, which is to multiply the derived {\HI} mass to optical light ratio ${\MHI /L}$ with the optical luminosity density \citep[e.g.][]{Delhaize:2013, Rhee:2013}. In the redshift range of our stacking experiment, the GAMA survey has high completeness in most of their surveyed optical bands, and a much larger sky coverage than the area investigated here. 
We focus on $r$-band data, because it provides one of the best luminosity density estimates, and has high target completeness and redshift success rates. To convert the measured {\HI} mass to {\HI} mass density ({\rhoHI}) we use:
\begin{equation} \label{eq:rho_HI_half}
{\rhoHI}^{'} = \dfrac{\left<{\MHI}\right>}{\left<L_r\right>} \rho_r .
\end{equation}

However, the H\,\textsc{i} mass-to-light ratio depends on luminosity and since, GAMA is a magnitude limited survey, {\rhoHI} will be potentially over-influenced by bright, relatively low ${\MHI /L}$ galaxies. This is straightforward to deal with by assuming a power-law relationship between the mass-to-light ratio and luminosity:
\begin{equation} \label{eq:power_law}
\dfrac{{\MHI}}{L_r} = 10^\kappa L_r^\beta .
\end{equation}

We also assume that the overall luminosity distribution is described by a Schechter function:
\begin{equation} \label{eq:schechter}
\phi(L)dL = \phi^\star\left(\frac{L}{L^\star}\right)^\alpha {\rm exp} \left(-\frac{L}{L^\star}\right)\frac{dL}{L} .
\end{equation}

A completeness correction factor can then be determined to adjust for the expected ratio between ${\MHI}/L_{r}$ for the total sample and the observed sample (see Appendix~A in \citet{Rhee:2013}):
\begin{equation} \label{eq:f_correct}
\begin{aligned}
&f=\dfrac{\left<{\MHI}\right>_{\rm all}/\left<L_r\right>_{\rm all}}{\left<{\MHI}\right>_{\rm obs}/\left<L_r\right>_{\rm obs}}\\
&\ =\dfrac{\int_0^\infty L^{\beta+1}\phi(L)dL}{\int_0^\infty L\phi(L)dL}\dfrac{\int_0^\infty LN(L)dL}{\int_0^\infty L^{\beta+1}N(L)dL} \\
&\ =\dfrac{{L^\star}^\beta\Gamma(2+\alpha+\beta)}{\Gamma(2+\alpha)}\dfrac{N}{\int_0^\infty L^{\beta+1}N(L)dL} ,
\end{aligned}
\end{equation}
where $\Gamma$ is the complete gamma function. We can then obtain an {\HI} density by applying this correction factor to Eq.~\ref{eq:rho_HI_half} as follows:
\begin{equation} \label{eq:rho_HI}
\rho_{\rm H\,\textsc{i}} = \rho^{'}_{\rm H\,\textsc{i}} f =\dfrac{\left<M_{\rm H\,\textsc{i}}\right>}{\left<L_r\right>} \rho_r f ,
\end{equation}
where $\left<{\MHI}\right>$ is the stacked {\HI} mass. The stacked $r$-band luminosity $\left<L_r\right>$ is calculated using the same weights as for $\left<{\MHI}\right>$, but choosing the median value of $\sigma_i$ as the noise weighting factor for each cubelet. We adopted $\beta = -0.4$ as used in \citet{Rhee:2013}.

\citet{Loveday:2012} studied the $u, g, r, i, z$ luminosity functions of GAMA. We adopt their $r$-band low-redshift ($z<0.1$) luminosity function, with Schechter  parameters from their table~3:
\begin{equation} \label{eq:loveday_paras}
\alpha=-1.26;\
M_{\star}=-21.50;\
\rho_r=1.225\times 10^8~{\Lsun}~{\rm Mpc}^{-3} ,
\end{equation}
where $M_{\star}$ corresponds to $L_{\star}=2.88\times 10^{10}~{\Lsun}$ \citep{Willmer:2018}.
We then calculate corresponding correction factors and cosmic {\HI} densities from Equations~\ref{eq:f_correct} and \ref{eq:rho_HI} for different values of the weighting parameter $\gamma$. The cosmic {\HI} density parameter ({\OHI}) is defined as:
\begin{equation} \label{eq:Omega_HI}
{\OHI} = \dfrac{\rhoHI} {\rho_{crit,0}}=\dfrac{8\pi G {\rhoHI}}{3H_0^2} ,
\end{equation}
where the critical density at $z=0$ is defined as $\rho_{crit,0}=1.36\times10^{11}~{\Msun}~{\rm Mpc}^{-3}$.

The results for {\OHI} are summarised in Table~\ref{tab:HI_table}. The uncertainties come from the jackknife re-sampling method described in Section~\ref{s:resultserrors}.

\begin{figure*}
    \includegraphics[width=\textwidth]{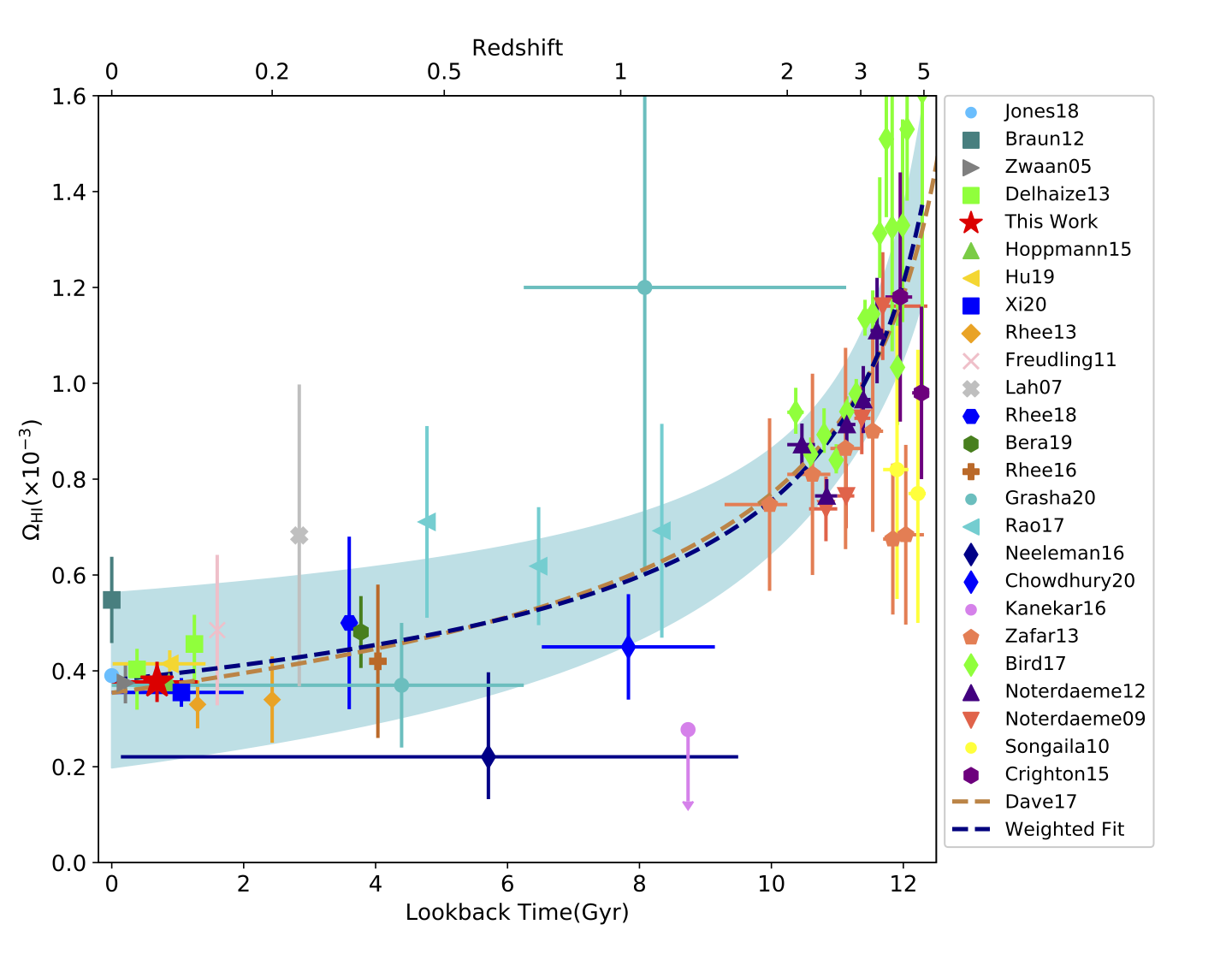}
    \caption{The cosmic neutral hydrogen density {\OHI} plotted as a function of redshift and lookback time. The sources for the measurements are noted in the right-hand legend, sorted by redshift. All measurements are corrected to the  cosmology used here. Similarly, DLA measurements which add other components (Helium, molecules) are corrected back to {\HI} only. \citet{Braun:2012} analysed the {\HI} absorption in nearby galaxies and applied an opacity correction; the \citet{Jones:2018} measurement is from the ALFALFA 100\% survey;  \citet{Delhaize:2013} has data points from two Parkes studies: HIPASS and a deep observation towards the south galactic pole; the result presented here is plotted as the red star; \citet{Freudling:2011}, \citet{Hoppmann:2015} and \citet{Xi:2020} are direct detection measurements from the Arecibo Ultra Deep Survey (AUDS); \citet{Rhee:2013} and \citet{Hu:2019} are both {\HI} stacking experiments using the WSRT; \citet{Zwaan:2005} is the HIPASS measurement; \citet{Lah:2007}, \citet{Rhee:2016} and \citet{Rhee:2018} use stacked GMRT 21-cm emission-line data; \citet{Bera:2019} is also a spectral stacking measurement using GMRT 21-cm emission data; the two \citet{Grasha:2020} measurements are from DLA data from the Green Bank Telescope (GBT); \citet{Rao:2017} is a DLA measurement based on a sample of MgII obsorbers; \citet{Neeleman:2016} is a DLA measurement based on HST archival UV data; \citet{Chowdhury:2020} is an {\HI} stacking experiment using GMRT; the \citet{Kanekar:2016} upper limit is from stacked GMRT {\HI} emission-line data for bright star forming galaxies; \citet{Zafar:2013} quotes DLA measurements from ESO UVES; \citet{Noterdaeme:2009}, \citet{Noterdaeme:2012}, \citet{Bird:2017} are DLA analyses using SDSS DR7, DR9 and DR12, respectively; \citet{Songaila:2010} provide DLA measurements from Keck data; \citet{Crighton:2015} quote results from a Gemini GMOS study of DLAs. The dashed light orange line is from the \citet{Dave:2017} model; the dashed navy blue line is a weighted linear fit to all the available measurements, with the shaded region representing the 95\% confidence region.}
    \label{fig:density_1}
\end{figure*}

With our preferred distance weighting index $\gamma=-1$, the stacked {\HI} mass is $(1.674\pm 0.183)\times 10^9 {\Msun}$, resulting in a stacked mass-to-light ratio of $(0.299\pm 0.033)  {\Msun}/{\Lsun}$. The derived cosmic density is $ (0.377\pm0.042)\times 10^{-3} $. The VLA has an excellent flux scale accuracy to 3\%-5\% \citep{Perley:2017a}. The errors reported are thus from jackknife analysis combined with an additional 5\% assumed as the fluxscale uncertainty. We plot our result in Fig.~\ref{fig:density_1}, together with other results. Our result is consistent with other work in this redshift range, but with high formal accuracy.

\subsection{Method 2: luminosity bins}

We can also directly measure $\left<{\MHI}\right>/\left<L_r\right>$ as a function of $\left<L\right>$ and extrapolate on to the universal galaxy luminosity function. We split our sample into six luminosity bins. With more than six, the stacked PSFs become less stable, and the S/N ratio becomes too small.  The luminosity range and number counts of galaxies and cubelets for the six bins is shown in Table~\ref{tab:Lbin_info}.

\begin{table}
	\centering
	\caption{Number of $z<0.1$ DINGO-VLA galaxies, and number of cubelets for six $r-$band luminosity bins.}
	\label{tab:Lbin_info}
	\begin{threeparttable}
        \begin{tabular}{cccccccc}
                \hline
               Bin & $N_{\rm gal}$ & $N_{\rm cubelets}$ & Luminosity Range \\
                       &                  &                          & $ (10^9 {\Lsun})$ \\
               \hline
               1 & 600 & 894 & 0.02 $\sim$ 0.86  \\
               2 & 600 & 895 & 0.86 $\sim$ 1.62  \\
               3 & 600 & 901 & 1.62 $\sim$ 2.76  \\
               4 & 600 & 901 & 2.76 $\sim$ 5.07  \\
               5 & 600 & 912 & 5.07 $\sim$ 11.44  \\
               6 & 622 & 939 & 11.45 $\sim$ 159.57       \\ \hline
        \end{tabular}
    \end{threeparttable}
\end{table}

For these six bins, we implement the above stacking and deconvolution procedure, and derive the stacked spectra. A constant frequency integration width of 3.0625~MHz is retained. The {\HI} density can then be measured without any correction factor as:
\begin{equation} \label{eq:rho_bins_half}
{\rhoHI}= \int_{L_{\rm min}}^\infty {\MHI}(L)\phi(L)dL ,
\end{equation}
where $\phi(L)$ is the standard Schechter function. To better extrapolate to lower masses, we can also assume a power law relation ${\MHI} /L\sim L^\beta$, and solve for $\kappa$ and $\beta$ in:
\begin{equation} \label{eq:linear_relation}
\log_{10}\dfrac{\left<{\MHI}\right>}{\left<L_r\right>} = \kappa + \beta\ \log_{10}\left<L_r\right> .
\end{equation}
The results are shown in Table~\ref{tab:Lbin_table}. It suggests that the slope of $\beta$ is steeper than the $-0.4$ used previously (more negative), and therefore that the contribution from optically-faint galaxies will be significant. It also means that the contribution of low-mass galaxies to the cosmic {\HI} density could be very significant if the slope of the HI mass function steepens, or even if it remains constant at low masses. For the time being, we only calculate the contribution of galaxies with $L_r>10^7$ {\Lsun}. By using $L_{\rm min}=10^7$ {\Lsun} in Eq.~\ref{eq:rho_bins_half} and combining with Equations~\ref{eq:schechter} and \ref{eq:linear_relation}, we calculate:
\begin{equation} \label{eq:rho_bins}
{\rhoHI}=10^\kappa{\phi^\star}^{\beta+1}\gamma_{up}\left(\alpha+\beta+2, \frac{L_{low}}{L_\star}\right) \Msun\ {\rm Mpc^{-3}},
\end{equation}
where $\gamma_{up}$ is the upper incomplete gamma function, defined as
\begin{equation} \label{eq:gamma_function}
\gamma_{up}(\xi,\theta) = \int_\theta ^\infty t^{\xi-1} e^{-t}dt .
\end{equation}
The results for ${\OHI}$ are shown in Table \ref{tab:Lbin_table}. The uncertainties come from jackknife re-sampling, with 10 jackknife re-samples combined with 5\% fluxscale error. The integrated and peak S/N are calculated in the same manner as in Table~\ref{tab:HI_table}. The derived values for {\OHI} are all slightly larger compared to method 1, but with excellent overall agreement. However, this method relies on successful parameter fit of $\kappa$ and $\beta$, which requires accurate {\MHI} measurements in the six stacking sub-samples. This makes it less certain than the first method. We formally prefer the $\gamma=-1$ value of ${\OHI}=(0.377\pm 0.042)\times 10^{-3}$ in Table~\ref{tab:HI_table}.


\begin{table*}
	\centering
	\caption{Results derived from stacking six luminosity sub-samples. Column 1 is weighting index $\gamma$; column 2 is the luminosity bin number; columns 3 to 5 are the weighting averaged redshift, \textit{r}-band luminosity and the derived {\HI} mass values, respectively; columns 6 and 7 are the integrated and peak S/N ratios for the extracted spectra, respectively; columns 8 and 9 are the parameter fits from Eq.~\ref{eq:linear_relation}}
	\label{tab:Lbin_table}


\begin{tabular}{ccccccccccccr}
\hline
 $\gamma$                                & Bin & $ \left \langle z \right \rangle$      & $\left \langle L_r \right \rangle$      & $\left \langle {\MHI} \right \rangle$ & Integrated SNR &  Peak SNR  & $\beta$                       & $\kappa$  & ${\OHI}$                    \\
                                      &        &            &    $ (10^9 {\Lsun})$  &  $(10^9 {\Msun})$ & &  &  &  &  $(10^{-3})$ \\
 (1) & (2) & (3) & (4) & (5) & (6) & (7) & (8) & (9) & (10) \\
\hline
\hline
{\multirow{6}{*}{0}}           & 1 & 0.0401 & 0.475  & 0.716 $\pm$ 0.148 &  8.76 &  8.04 &   \multirow{6}{*}{$-0.528$} & \multirow{6}{*}{4.735}  & \multirow{6}{*}{0.465}  \\
                                        & 2 & 0.0635 & 1.237  & 0.621 $\pm$ 0.316 &  3.55 &  4.37 &                       &                      &  \\
                                        & 3 & 0.0690 & 2.128  & 1.646 $\pm$ 0.481 &  8.27 &  5.16 &                       &                      &  \\
                                        & 4 & 0.0698 & 3.804  & 2.831 $\pm$ 0.506 &  11.94 &  5.16 &                       &                       & \\
                                        & 5 & 0.0690 & 7.696  & 2.959 $\pm$ 0.645 &  12.63 &  5.67 &                       &                      &  \\
                                        & 6 & 0.0685 & 25.508 & 3.408 $\pm$ 0.762 &  16.67 &  4.97 &                       &                      &  \\
\hline
\multirow{6}{*}{$-1$}            & 1 & 0.0309 & 0.383  & 0.583 $\pm$ 0.084 & 11.73 &  10.59 & \multirow{6}{*}{$-0.529$} & \multirow{6}{*}{4.740} & \multirow{6}{*}{0.455}\\
                                        & 2 & 0.0571 & 1.219  & 0.680 $\pm$ 0.274 &  4.84 &  5.17 &                       &                       & \\
                                        & 3 & 0.0607 & 2.122  & 1.648 $\pm$ 0.405 &  10.71 &  6.68 &                       &                        &\\
                                        & 4 & 0.0615 & 3.813  & 2.603 $\pm$ 0.430 &  14.03 &  6.11 &                       &                       & \\
                                        & 5 & 0.0626 & 7.746  & 2.908 $\pm$ 0.526 &  15.44 &  6.83 &                       &                     &   \\
                                        & 6 & 0.0624 & 25.469 & 3.246 $\pm$ 0.637 &  19.36 &  5.79 &                       &                     &   \\
\hline
\multirow{6}{*}{$-2$}            & 1 & 0.0214 & 0.283  & 0.423 $\pm$ 0.045 & 17.33 &  15.48 & \multirow{6}{*}{$-0.516$} & \multirow{6}{*}{4.586} & \multirow{6}{*}{0.409}\\
                                        & 2 & 0.0471 & 1.191  & 0.726 $\pm$ 0.222 &  6.75 &  6.41 &                       &                      &  \\
                                        & 3 & 0.0465 & 2.107  & 1.566 $\pm$ 0.307 &  14.59 &  9.14 &                       &                      &  \\
                                        & 4 & 0.0475 & 3.829  & 2.162 $\pm$ 0.339 & 18.36 &  8.40 &                        &                      &  \\
                                        & 5 & 0.0538 & 7.828  & 2.825 $\pm$ 0.413 & 19.35 &  8.30 &                        &                      &  \\
                                        & 6 & 0.0546 & 25.160 & 3.030 $\pm$ 0.523 &  22.44 &  6.79 &                       &                      &  \\
\hline
\multirow{6}{*}{$-3$}            & 1 & 0.0144 & 0.201  & 0.284 $\pm$ 0.049 & 22.60 &  20.54 & \multirow{6}{*}{$-0.497$} & \multirow{6}{*}{4.340} & \multirow{6}{*}{0.339}\\
                                        & 2 & 0.0339 & 1.150  & 0.736 $\pm$ 0.171 &  9.54 &  8.30 &                       &                     &   \\
                                        & 3 & 0.0298 & 2.081  & 1.316 $\pm$ 0.226 &  18.52 &  12.28 &                       &                     &   \\
                                        & 4 & 0.0303 & 3.865  & 1.536 $\pm$ 0.380 & 27.19 &  14.10 &                        &                     &   \\
                                        & 5 & 0.0433 & 7.935  & 2.667 $\pm$ 0.341 & 22.97 &  9.37 &                        &                     &   \\
                                        & 6 & 0.0451 & 24.635 & 2.732 $\pm$ 0.449 & 25.73 &  7.97 &                        &                    &   \\
\hline
\end{tabular}

\end{table*}

\section{Summary and Conclusion}\label{section:summary}


We have applied a new \textit{Cubelet Stacking} technique to the $z<0.1$ data from the DINGO-VLA survey using a pipeline  developed specifically to cope with large survey data. As the {\HI} signals from the sample galaxies are weak and potentially resolved by the interferometer, the traditional stacking method does not work particularly well, and leads to large flux errors.  We therefore stacked image and PSF cubes and deconvolved the resultant detections. This is not possible in normal {\HI} spectral stacking. Using this method, 5442 cubelets from 3622 galaxies are stacked, then the resultant cube was deconvolved. As shown in Paper I, a deep 1-$\sigma$ threshold {\sc clean} and a aperture radius of $R=20^{''}$  results in the most accurate stacked spectrum. We investigate different sample weighting schemes ($w=\sigma^{-2}D^\gamma$) and and find that $\gamma=-1$ is the best compromise, giving good S/N ratio results for the stacked data and low cosmic variance. We make a $30\sigma$ measurement of the stacked {\HI} mass for our sample, $\left<\MHI\right>=(1.674 \pm  0.183)\times10^9~{\Msun}$. The corresponding cosmic density is ${\OHI}=(0.377 \pm 0.042)\times 10^{-3}$ at $z\sim 0.051$. We also directly measure the cosmic {\HI} density in six independent luminosity bins and obtain results in good agreement. All data points in Fig. \ref{fig:density_1} have been adjusted to the 737 cosmology framework (i.e., $\rm H_0=70\ km\ s^{-1}Mpc^{-1}$, $\rm \Omega_{M}$=0.3, $\rm \Omega_\Lambda$=0.7). We should note that we assume that the {\HI} is optical thin, and that self-absorption is negligible. Previous work on nearby galaxies suggests that corrections of the order of 10s of per cent may be required \citep{Stanimirovic:1999, Liu:2019}. Systematic errors may also arise from incomplete optical spectroscopy. However, the G09 data being used in this work has an overall completeness of 98.48\% at $r<19.8$, so incompleteness down to this limit should not affect our measurements. 
Finally, the DINGO-VLA data analysed in this work occupies $\sim$38 deg$^2$ equatorial area with redshift spans from 0.002 to 0.1. The cosmic variance in this region is as large as 27.05\%.\footnote{From https://cosmocalc.icrar.org.} The $r$-band luminosity density used in this paper is generated from multiple GAMA regions, and the overall cosmic variance in GAMA is less than 10\% \citep{Driver:2010a}. But still, this is the an important uncertainty source in this work.

Including our work, measurements of cosmic HI density beneath $z\sim 0.4$ largely agree with each other (see Fig.~\ref{fig:density_1}). These results indicate that {\OHI} has not evolved significantly of the past $\sim$4 Gyr. However, at higher redshifts, {\OHI} shows a significant increase, especially for those measurements at $z>2$ using DLAs. We conduct a simple linear fit in redshift of all these measurements, weighted by their uncertainties. We show the fitting result by the dashed blue line, with the shaded region indicating the 95\% confidence interval. 
However, it is known that theory cannot yet fully explain this evolution, possibly due to incomplete understanding of the complex physical situations which determine the boundaries between different gas phases. Predictions from semi-analytical models \citep[such as][]{Lagos:2011,Lagos:2014,Lagos:2018,Popping:2014,Kim:2015,Power:2010} often agree well with observations at low redshift, but at $z>0.4$ they show considerable discrepancies. \citet{Dave:2017} uses the hydrodynamical simulation {\sc mufasa} and derives a evolution trend of ${\OHI}=10^{-3.45}(1+z)^{0.74}$. This model is in excellent agreement with observation results shown in Fig \ref{fig:density_1}. We show this curve as the dashed orange line.

The {\OHI} measurements based on 21~cm emission lines mostly cluster at $z<0.2$. Our future work will examine the higher redshift portion of the DINGO-VLA data, which will provide a measurement of stacked {\HI} content to $z\sim 0.3$, and contribute to the robust study of $\OHI$ over longer cosmological baselines. Moreover, by stacking the input GAMA galaxy sample into bins of stellar mass, colour and environment (group, cluster etc), the DINGO-VLA dataset will enable studies of gas scaling relations, gas depletion and the environmental effects on galaxy evolution. This developed cubelet stacking methodology will also be particularly useful for {\HI} size related studies. Our future work will examine the higher redshift portion of the DINGO-VLA data (see Fig. \ref{fig:density_1}) which will allow these studies to examine the evolution of these quantities over cosmic time.

\section*{Acknowledgements}
\addcontentsline{toc}{section}{Acknowledgements}




Parts of this research were supported by the Australian Research Council Centres of Excellence for All Sky Astrophysics in 3 Dimensions (ASTRO 3D) and All-sky Astrophysics (CAASTRO), through project numbers CE170100013 and CE110001020, respectively. Chen acknowledges Dr. Barbara Catinella, Dr. Jonghwan Rhee and Dr. Wenkai Hu for helpful discussions. The National Radio Astronomy Observatory is a facility of the National Science Foundation operated under cooperative agreement by Associated Universities, Inc.

\section*{Data Availability}

The raw DINGO-VLA data is available from the on-line NRAO data archive\footnote{science.nrao.edu/facilities/vla/archive} under projects VLA/14B-315 and VLA/16A-341. The optical data for the GAMA G09 field is available from the project website\footnote{www.gama-survey.org}. The derived data supporting the findings of this study are available from the corresponding author on reasonable request.

\bibliographystyle{mnras}
\bibliography{ref}


\bsp	
\label{lastpage}
\end{document}